\pdfoutput=1
\documentclass[aps,prl,superscriptaddress,twocolumn,twoside,dvipsnames]{revtex4-2}

	\usepackage[T1]{fontenc}
	\usepackage{amsmath}
	\usepackage{amssymb}
	\usepackage{mathrsfs}
	\usepackage{dsfont}
	\usepackage{verbatim}
	\usepackage{tikz}
	\usepackage{xcolor}
	\usepackage[
      colorlinks=true,
      linkcolor=blue,
      urlcolor=cyan,
      filecolor=magenta,
      citecolor=red,
      pdfstartview=FitV,
      pdftitle={Non-Lorentzian Chaos and Cosmological Holography},
        pdfauthor={Arjun Bagchi, Shankadeep Chakrabortty, Daniel Grumiller, Bharathkumar Radhakrishnan, Max Riegler, Aditya Sinha},
        pdfsubject={Non-Lorentzian Chaos and Cosmological Holography},
        pdfkeywords={Non-Lorentzian, Chaos, Cosmological Holography},
        bookmarksopen=true
      ]{hyperref}

	\newcommand{\htp}{\tau_{+}}
	
	\newcommand{\be}[1]{\begin{equation}\label{#1} }
	\newcommand{\ee}{\end{equation}}
	\newcommand{\bea}[1]{\begin{eqnarray}\label{#1} }
	\newcommand{\eea}{\end{eqnarray}}
	
	\newcommand{\refb}[1]{(\ref{#1})}
    \newcommand*{\scri}{\ensuremath{\mathscr{I}}}

	\renewcommand{\>}{\rangle}

	\newcommand{\unity}{\mathds{1}}

	\def\pb{\frac{2\pi}{\beta}}
	\def\cA{\mathcal{A}}
	\def\cF{\mathcal{F}}
	\def\cO{\mathcal{O}}
	\def\nn{\nonumber}
	\def\bec{\begin{center}}
	\def\eec{\end{center}}
	\def\beq{\begin{equation}}
	\def\eeq{\end{equation}}
	\def\bseq{\begin{subequations}}
	\def\eseq{\end{subequations}}
	\def\bea{\begin{eqnarray}}
	\def\eea{\end{eqnarray}}

    \newcommand{\mass}{M}
    \newcommand{\angmom}{J}
    \newcommand{\massp}{m}
    \newcommand{\jp}{j}	

\begin{document}
	\newcommand{\mytitle}{Non-Lorentzian Chaos and Cosmological Holography}
	
	    \title{\mytitle}

        \author{Arjun Bagchi}
        \email[]{abagchi@iitk.ac.in}

        \affiliation{Indian Institute of Technology, Kanpur, 208016 India}
        
        \author{Shankhadeep Chakrabortty}
        \email[]{s.chakrabortty@iitrpr.ac.in}

        \affiliation{Indian Institute of Technology, Rupnagar, Punjab 140001, India}
        
        \author{Daniel Grumiller}
        \email[]{grumil@hep.itp.tuwien.ac.at}

        \affiliation{Institute for Theoretical Physics, TU Wien,
Wiedner Hauptstrasse 8–10/136, A-1040 Vienna, Austria}

        \author{Bharathkumar Radhakrishnan}
        \email[]{r.bharathkumar@outlook.com}

        \affiliation{Indian Institute of Science Education and Research, Mohali, Punjab 140306, India}
        \affiliation{International Centre for Theoretical Sciences, Bangalore, Karnataka 560089, India}

\author{Max Riegler}
        \email[]{mriegler@fas.harvard.edu}

        \affiliation{Center for the Fundamental Laws of Nature,
Harvard University, MA 02138, USA}
        
\author{Aditya Sinha}
        \email[]{asinha5@andrew.cmu.edu}

        \affiliation{Carnegie Mellon University, Pittsburgh, PA 15213, USA}
        \date{\today}

    \begin{abstract}
We study chaos in non-Lorentzian field theories, specifically Galilean and Carrollian conformal field theories in two dimensions. In a large central charge limit, we find that the Lyapunov exponent saturates the bound on chaos, conjectured originally for relativistic field theories. We recover the same Lyapunov exponent holographically by a shock-wave calculation in three-dimensional flat space cosmologies, providing further evidence for flat space holography.
    \end{abstract}

    \maketitle
    
	\flushbottom

{\em{Introduction.}}
As aptly put by Edward Lorenz, a dynamical system becomes chaotic when the present determines the future, but the approximate present does not. In other words, the dynamics of the system is strongly sensitive to initial conditions. Colloquially, this is often described by flapping butterfly wings in Brazil, causing a tornado in Texas \cite{Lorenz:1972}. In this work, we discuss aspects of chaos in unconventional quantum field theories (QFTs) and their holographic manifestation in cosmological spacetimes.

The QFTs we consider are non-Lorentzian conformal field theories (CFTs) in two dimensions (2d). We begin our analysis with Galilean CFTs (GCFTs) \cite{Bagchi:2009my}, where the speed of light goes to infinity \cite{Bacry:1968zf}. GCFTs are natural analogues of relativistic CFTs \cite{diFrancesco}. They are expected to arise as fixed points in renormalization group flows in the parameter space of Galilean QFTs. GCFTs appear in the non-relativistic limit of all known relativistic conformally invariant QFTs, e.g.~massless scalars and fermions in all dimensions, sourceless electrodynamics, and pure non-Abelian gauge theories in four dimensions \cite{Bagchi:2017yvj}. GCFTs are governed by the Galilean conformal algebra (GCA), which in 2d reads  \cite{Bagchi:2009pe}
\begin{subequations}
\label{bms}
\begin{align}
[L_n, L_m] &= (n-m) L_{n+m} + \frac{c_L}{12}\big(n^3-n\big)\delta_{n+m,0} \\ 
[L_n, M_m] &= (n-m) M_{n+m} + \frac{c_M}{12}\big(n^3-n\big)\delta_{n+m,0} \\ 
[M_n, M_m] &= 0\qquad\qquad\qquad\qquad\qquad\quad n,m\in\mathbb{Z}\,. 
\end{align}
\end{subequations}
The $c$-numbers $c_L$ and $c_M$ are the central charges. We focus on the case $c_L=0$ and large $c_M$, since this will be most relevant for holographic applications of our results. With methods mirroring recent advances in relativistic 2d CFTs \cite{Roberts:2014ifa,Fitzpatrick:2015zha,Perlmutter:2016pkf} we compute the diagnostics of chaos in GCFTs, specifically, the Lyapunov exponent, which characterizes the divergence of nearby trajectories in phase space.

We are also interested in another class of non-Lorentzian CFTs, viz.~Carrollian CFTs (CCFTs). In Carrollian theories, the speed of light goes to zero, the opposite of the Galilean limit \cite{Bacry:1968zf, Bagchi:2012cy, Duval:2014uoa, Bagchi:2016bcd}. Carrollian QFTs encompass any QFT defined on a null manifold \cite{Bagchi:2019clu}. Interestingly, in $d=2$, the Galilean and Carrollian groups are isomorphic. This isomorphism extends to their conformal versions \cite{Bagchi:2010zz}. Thus, 2d CCFTs are also governed by the algebra \refb{bms}. Hence our answers, constructed for 2d GCFTs, are also valid for 2d CCFTs, albeit with an interchange of the spatial and temporal directions, and correspondingly different physical interpretations.

Finally, we use the holographic correspondence between 3d flat space and 2d CCFTs \cite{Bagchi:2010zz, Bagchi:2012cy} (elaborated on later) to holographically reproduce our field theory answers. Thermal states in the CCFT are dual to cosmological solutions called `Flat Space Cosmologies' (FSC) \cite{Cornalba:2002fi,Cornalba:2003kd}. A calculation similar in spirit to the corresponding anti-de~Sitter (AdS) black hole calculation with a shock-wave \cite{Shenker:2013pqa,Shenker:2013yza} recovers the Lyapunov exponent that we find on the field theory side, giving further support to the proposed holographic correspondence in flat space. The main result of the present work is that both on the field theory and the gravity sides the Lyapunov exponent
\begin{equation}
\lambda_L = \frac{2\pi}{\beta}
\label{eq:angelinajolie}    
\end{equation}
saturates the conjectured bound on chaos \cite{Maldacena:2015waa} ($\beta$ is the inverse temperature).

{\em{Galilean CFT basics.}}  
We begin with some representation theory aspects of 2d GCFTs. The states of the theory are labeled by \cite{Bagchi:2009pe}
\be{eq:weights}
L_0|\Phi\> = \Delta_\Phi |\Phi\>\,, \qquad  M_0|\Phi\> = \xi_\Phi |\Phi\>\,.
\ee
The action of $L_n, M_n$ for $n>0$ on a state $|\Phi\>$ lowers the weight $\Delta_\Phi$. Hence in analogy with 2d CFTs, there is a notion of primary states $|\Phi\>_p$ satisfying 
\be{}
L_n|\Phi\>_p =  M_n|\Phi\>_p = 0,\qquad\forall n>0\,.
\ee
The rest of the module is built by acting with raising operators $L_{-n}, M_{-n}$ for $n>0$ on a given primary state. Like in relativistic CFTs, there is also a state-operator correspondence $\Phi(0,\,0) \leftrightarrow|\Phi\>$. Below we exclusively use operators. 

Chaos in classical systems can be diagnosed via the sensitivity to initial conditions using the Poisson bracket $\{x(t),p(0)\}=\frac{\partial x(t)}{\partial x(0)}$, which can grow as a sum of exponentials in $t$. The exponents therein are called Lyapunov exponents. The analogous quantity for a quantum system in a state that is described by a density matrix $\rho$ is $-\textnormal{Tr}(\rho[x(t),p(0)])$ \cite{Larkin:1969aa}. Random phase cancellations that thermalize this quantity too soon are avoided by considering the square of the commutator. To study chaos of a thermal system in equilibrium at inverse temperature $\beta$ we thus consider the quantity \cite{Shenker:2013yza,Roberts:2014isa,Shenker:2014cwa,Kitaev:2014aa}
    \begin{equation}\label{eq:SqaredCommutator}
        C(t)=-\langle[W(t),V(0)]^2\rangle_\beta\,,
    \end{equation}
which can be written in terms of time-ordered and out-of-time-ordered correlation functions (OTOCs). In a generic quantum many-body system the former approach a constant after the relaxation time $\beta$ while the OTOCs in contrast start at a large value and decrease over time, resulting in an increase of $C(t)$. Thus, one can use OTOCs to diagnose chaos \cite{Haehl:2017qfl}. To find chaotic behavior in a generic 2d GCFT we study the late time behavior of the following OTOC of pairs of local primary operators in a thermal state
\begin{equation}\label{eq:ThermalOTOC}
\frac{\langle V^\dag(0) W^\dag(t) V(0) W(t) \rangle_\beta}{\langle V^\dag(0) V(0)\rangle_\beta\langle W^\dag(t)W(t)\rangle_\beta}\,,
\end{equation}
and look for exponential decay of this OTOC. It is evaluated by mapping the 2d GCFT from the plane to a cylinder of radius $\beta$. We work in the approximation $c_M /\xi_n \gg 1$~\footnote{%
The approximation of large central charge corresponds to the classical approximation on the gravity side (small Newton constant) \cite{Barnich:2006av}, whenever there is a gravity dual. Therefore, this approximation is not only convenient, it is also pertinent to the holographic calculation in the second half of our work.
}, where $\xi_n=\xi_V, \xi_W$ are the weights defined in \eqref{eq:weights}. This allows the use of closed-form expressions of Galilean conformal blocks \cite{Bagchi:2016geg,Bagchi:2017cpu,Hijano:2017eii,Hijano:2018nhq,Merbis:2019wgk}.
        
{\em{Chaotic correlators in GCFT.}} We start with a 2d GCFT on a complex plane with coordinates $(u, v)$. Consider two local scalar primary operators $V$ and $W$ with weights $(\Delta_V,\xi_V)$ and $(\Delta_W,\xi_W)$, respectively. Symmetry restricts the vacuum 4-point function \cite{Bagchi:2009pe}
\beq\label{eq:Euclidean-amplitude}
\frac{\langle V^\dag(u_1,v_1) V(u_2,v_2) W^\dag(u_3,v_3) W(u_4,v_4) \rangle}{\langle V^\dag(u_1,v_1) V(u_2,v_2)\rangle \langle W^\dag(u_3,v_3) W(u_4,v_4) \rangle } = \cA
\eeq
to be a function $\cA=\cA(\chi,\zeta)$	of the GCFT cross-ratios
\beq\label{eq:lalapetz}
\chi =\frac{u_{12}u_{34}}{u_{13}u_{24}}\,, \qquad \frac{\zeta}{\chi} = \frac{v_{12}}{u_{12}} + \frac{v_{34}}{u_{34}} - \frac{v_{13}}{u_{13}} - \frac{v_{24}}{u_{24}}\,, 
\eeq
where $\mbox{$u_{nm}:=u_n-u_m$}$ and $\mbox{$v_{nm}:=v_n-v_m$}$. The amplitude $\cal A$ is invariant under a map to the (thermal) cylinder $u=\exp{(\pb\tau)}$, $v=\pb \sigma \exp{(\pb\tau)}$, with a spacelike coordinate $-\infty<\sigma<\infty$ and complexified time $\tau=t_R+it_E$ where $t_R$ and $t_E$ denote real and Euclidean time, and $\tau$ satisfies $\tau\sim\tau+i\beta$. 
        
The OTOC \eqref{eq:ThermalOTOC} is obtained by an analytic continuation from the Euclidean version of \eqref{eq:Euclidean-amplitude} involving three steps. First, one separates all operators in Euclidean time $t_{E_n}=\epsilon_n$ such that $\epsilon_1<\epsilon_3<\epsilon_2<\epsilon_4$. This preserves the real-time ordering of the operators in \eqref{eq:ThermalOTOC}. Second, the real-time is evolved until the desired values of $t_R$. Finally, one takes the $\epsilon_n$ to zero. We place $V$ and $V^\dag$ at $\sigma=t_R=0$ and $W$ as well as $W^\dag$ at $\sigma=x$, $t_R=t$. This leads to $(u_n,v_n)=(\exp{[\pb i\epsilon_n]},0)$ for $n\in\{1,2\}$ and $(u_n,v_n)=(\exp{[\pb(t+i\epsilon_n)]},\pb x\exp{[\pb(t+i\epsilon_n)]})$ for $n\in\{3,4\}$.

A useful choice of operator positions along the thermal circle is to place them in diametrically opposite pairs, i.e., $\epsilon_2=\epsilon_1+\beta/2$ and $\epsilon_4=\epsilon_3+\beta/2$. We set $\epsilon_1=0$ without loss of generality and define the angular displacement of both operator pairs as $\theta=\pb\epsilon_3$, which satisfies $0<\theta<\pi$, maintaining the required operator ordering. The cross-ratios $\chi$ and $\zeta$ simplify to
		    \beq
		    \label{eq:x}
				\frac{1}{\chi}  = - \sinh^2\Big(\frac{\pi t}{\beta}+\frac{i \theta}{2}\Big),\, \frac{\zeta}{\chi} = -\frac{2\pi x}{\beta}\coth\Big(\frac{\pi t}{\beta}+\frac{i \theta}{2}\Big).
			\eeq
The first equality \eqref{eq:x} shows that the cross-ratio $\chi$ encircles counter-clockwise the point $\chi=1$, which will turn out to be a branch point of the amplitude $\mathcal{A}$. 
The variable $\zeta$ follows a closed contour that does not enclose singularities or branch cuts of $\mathcal{A}$.

{\em{GCA blocks and Regge limits.}} For large values of $c_M/\xi_n$ the amplitude $\cA(\chi,\zeta)$ given by \eqref{eq:Euclidean-amplitude} can be written in terms of $s$-channel global GCA blocks \cite{Bagchi:2016geg,Bagchi:2017cpu} and 3-point coefficients $C^p_{nm}$
			\begin{align}
				    & \cA(\chi,\zeta) = \sum_p C^p_{VV}C^p_{WW} ~\mathfrak{g}^{VV}_{WW}(p|\chi,\zeta),	\quad \text{with} \label{eq:g1234} \\
					& \mathfrak{g}^{VV}_{WW} =  4^{\Delta_p-1} (1-\chi)^{\frac{1}{3}\left(\Delta_V+\Delta_W-\frac{3}{2}\right)}\chi^{\Delta_p+\frac{4}{3}\left(\Delta_W-2\Delta_V\right)} \nonumber \\ 
					& e^{\frac{\zeta}{\chi}\left(\frac{4}{3}\left(2\xi_V-\xi_W\right)-\frac{\xi_p}{\sqrt{1-\chi}}\right)+\frac{\zeta\left(\xi_V+\xi_W\right)}{3\left(1-\chi\right)}}( 1 + \sqrt{1-\chi})^{2-2\Delta_p} \nonumber
				\end{align}
Note that $\mathfrak{g}^{VV}_{WW}(p|\chi,\zeta)$ exhibits a branch cut along $\chi\in(1,\infty)$.
                
For large, negative values of $t$ the contribution from the identity channel ($p=0$) dominates since $\chi\approx0$ and $\zeta\approx0$. At late times $t\gg\beta$ one has again $\chi\approx0$ and $\zeta\approx0$. However, $\chi$ crosses the branch cut along $\chi\in(1,\infty)$ leading to a non-trivial monodromy of $\mathfrak{g}^{VV}_{WW}(p|\chi,\zeta)$. This process is equivalent to taking the GCA Regge limit \cite{Perlmutter:2016pkf}
	        \beq\label{eq:regge-limit}
				(1-\chi) \rightarrow e^{2\pi i}(1-\chi), \quad \chi,\zeta \rightarrow 0, \quad \frac{\zeta}{\chi} = \text{ const.}
			\eeq
In the limit \eqref{eq:regge-limit} the cross-ratios \eqref{eq:x} simplify to
				\beq
					\chi = - \exp\Big( - \pb \,t\Big)\epsilon_{12}\epsilon^{*}_{34}\,,\qquad\frac{\zeta}{\chi} = -\pb\, x\,,
				\eeq
with $\epsilon_{nm}:=i\,[\exp(\pb i\,\epsilon_n)-\exp(\pb i\,\epsilon_m)]$. The leading behavior of the amplitude \eqref{eq:g1234} is then given by
				\beq
					\mathfrak{g}^{\text{Regge}}_{V,W}(p|\chi,\eta) = N(p,V,W)\chi^{2-\Delta_p'} e^{-\pb x\,\xi_p'}\,,
				\eeq
where $N(p,V,W)=16^{\Delta_p-1}\exp{\big(\frac{\pi i}{3}[2(\Delta_V+\Delta_W)-3]\big)}$ and  $\Delta_p'=\Delta_p-\frac{4}{3}(\Delta_W-2\Delta_V)$, and similarly for $\xi_p'$. In terms of $x$ and $t$, the above expression is
				\beq\label{eq:firstmain}
					\frac{\mathfrak{g}^{\text{Regge}}_{V,W}(p|x,t)}{N(p,V,W)} = e^{-\pb t(2-\Delta_p') -\pb x \,\xi_p' } (\epsilon_{12}\epsilon^{*}_{34})^{2-\Delta_p'}\,.
				\eeq
Equation \eqref{eq:firstmain} is our first key result. 
		
{\em{Lyapunov exponent in GCFT.}} For large $c_M/\xi_n$ and at late times $t\gg\beta$ (but $t\ll t_\ast$, see below) the OTOC \eqref{eq:Euclidean-amplitude} is dominated by the identity block \cite{Merbis:2019wgk}
			\beq
				\cA_{\unity} \sim 1 + \frac{2}{c_M} (\Delta_V \xi_W + \Delta_W \xi_V  + \xi_V \xi_W \zeta \partial_\chi)\cF (\chi)\,,
			\eeq
where $\cF(\chi)=\chi^2\,_2F_1(2,2,4,\chi)=(6-12/\chi)\log(1-\chi)-12$. Applying the limit, we find
			\bea\label{eq:IdentityBlockLypunovDecay}
				& \cA_{\unity}(x,t) \sim 1 + h(x) \exp{\Big( \pb(t-t_*) \Big)}\,, \quad \text{with} \nn \\
				& h(x) = \frac{48\pi i}{\epsilon_{12}\epsilon^{*}_{34}}\Big[ (\Delta_V \xi_W + \Delta_W \xi_V) + \xi_V \xi_W \frac{2\pi}{\beta} x\Big]\,. 
			\eea
We define the scrambling time $t_\ast$ in \eqref{eq:IdentityBlockLypunovDecay} as the time-scale where the global block expansions fails~\footnote{%
Since $c_M$ has a physical dimension a more precise definition of scrambling time is $t_\ast\sim\frac{\beta}{2\pi}\log\frac{c_M}{\xi_n}$, where $\xi_n$ is either $\xi_V$ or $\xi_W$, depending on which weight is bigger. However, since we assume the weights to be finite and $c_M/\xi_n$ to be large, the quantity $c_M$ must be large and hence dominates the scrambling time.}: 
				\beq
					t_\ast \sim \frac{\beta}{2\pi}\,\log c_M\,.
				\eeq
From the exponential behavior of the OTOC \eqref{eq:IdentityBlockLypunovDecay} we read off the main result of our field theory calculations, the Lyapunov exponent for 2d GCFTs~\footnote{%
This result can be extended to the case where $c_L\neq 0$ since a non-zero $c_L$ only affects the $\cO(1/c_M^2)$ terms.}
			\beq\label{eq:whydidthishavenolabel}
			\lambda_L=\pb.
			\eeq
We conclude the GCFT discussion with a few technical remarks. Crossing the branch cut and passing to the second sheet of $\mathcal{A}$ is essential for obtaining a non-trivial result for the OTOC exhibiting chaotic behavior. The sign in front of $h(x)$ in \eqref{eq:IdentityBlockLypunovDecay} ensures that the magnitude of $\cA_{\unity}$ decreases in time for any choice of $\epsilon_n$ that is consistent with the real-time ordering of the operators in \eqref{eq:ThermalOTOC} and it does so in an exponential manner, signaling the onset of chaos in a 2d GCFT.

To obtain the corresponding Carrollian results relevant for flat space holography, one simply replaces $t \leftrightarrow x$. Thus, in stark contrast to CFTs, it is not an OTOC that exhibits chaotic behavior in a CCFT but rather an out-of-\emph{space}-ordered correlation function. The expression for the Lyapunov exponent stays the same as above. 

In the remainder of this work, we show how to reproduce the Lyapunov exponent \eqref{eq:whydidthishavenolabel} using holographic methods in flat space. We shall see that the interchange of space and time required for Carrollian results manifests itself on the gravity side as a change from black hole physics in standard AdS/CFT to a cosmological set-up in flat holography. 

{\em{Holography and Flat Space Cosmology.}} Although holography has been principally and successfully explored using the AdS/CFT correspondence \cite{Maldacena:1997re}, there are recent efforts generalising beyond AdS and specifically to 3d asymptotically flat spacetimes \cite{Bagchi:2010zz, Bagchi:2012cy, Barnich:2012aw, Bagchi:2012yk, Bagchi:2012xr, Barnich:2012xq,Barnich:2012rz,Bagchi:2013lma,Afshar:2013vka,Gonzalez:2013oaa,Fareghbal:2013ifa,Detournay:2014fva,Bagchi:2014iea,Bagchi:2015wna,Hartong:2015usd}. The asymptotic symmetry group (ASG) dictates the symmetries at the boundary of a spacetime. A natural recipe for holography is to declare this to be the symmetry governing the dual field theory. In asymptotically flat spacetimes, the ASG at null infinity is the Bondi-Metzner-Sachs (BMS) group. Following the above recipe, the 2d field theory duals of 3d flat space are governed by the BMS$_3$ algebra, which turns out to be \refb{bms} again, with $c_L=0$ and $c_M=3/G$, where $G$ is the 3d Newton constant \cite{Barnich:2006av}. In hindsight, this is not surprising, as the flat space limit of AdS translates to a Carrollian limit on the boundary CFT \cite{Bagchi:2012cy}. This means we can check our field theory results in a flat space holographic context. 

The zero modes of the most general solutions compatible with boundary conditions that generate the symmetries \refb{bms} at null infinity are FSC solutions \cite{Cornalba:2002fi,Cornalba:2003kd}. They describe toy model universes with a contracting and an expanding phase given by the locally Ricci-flat metric
	\begin{equation}\label{FSCmetric}
	    ds^2 = - \frac{d\tau^2}{f(\tau)} + f(\tau)dx^2 + \tau^2 \left(d\phi - N_{\phi}(\tau) dx\right)^2\,,
	\end{equation}
where $f(\tau)=\htp^2(\tau^2-\tau_{0}^2)/\tau^2$ and $N_{\phi}(\tau)=\tau_0\htp/\tau^2$. This geometry has a cosmological horizon at $\tau=\tau_{0}$, such that $\tau$ is timelike while $x,\phi$ are spacelike for all $\tau>\tau_0$. The parameters $\tau_+=\sqrt{\massp}$ and $\tau_0=|\jp|/\sqrt{\massp}$ determine the FSC mass $\mass=\frac{\massp}{8G}$ and angular momentum $\angmom=\frac{\jp}{4G}$.   

In 3d flat holography, the FSCs \eqref{FSCmetric} are dual to thermal states in the 2d CCFT discussed above. This is analogous to the AdS/CFT relation between BTZ black holes \cite{Banados:1992wn,Banados:1992gq} and thermal states in a 2d CFT.  Below, inspired by corresponding AdS shock-wave calculations \cite{Shenker:2013pqa,Roberts:2014isa,Maldacena:2015waa,Engelsoy:2016xyb,Grumiller:2020fbb}, we perform an analysis of a shock-wave in the FSC geometry to compute the Lyapunov exponent holographically, validating our field theory result \eqref{eq:whydidthishavenolabel}. 

{\em Null geodesics on FSC backgrounds.} A general null geodesic moving on the background \eqref{FSCmetric} satisfies 
                \begin{equation}\label{nullGeodFSC}
                    - \frac{\dot{\tau}^2}{f(\tau)} + f(\tau)\dot{x}^2 + \tau^2(\dot{\phi} - N_{\phi}(\tau) \dot{x})^2 = 0\,,
                \end{equation}
where dots denote derivatives with respect to an affine parameter. The first integrals of the geodesic equations
                \beq\label{eqFSC}
                    \massp\dot{x} - \jp\dot{\phi} = P\,, \qquad\tau^2 \dot{\phi} - \jp \dot{x} = L \,,
                \eeq
yield constants of motion $P$ and $L$ corresponding to the Killing vectors $\partial_x$ and $\partial_\phi$, respectively. For the sake of simplicity, we set $L=0$~\footnote{The case $L\neq0$ yields essentially the same results and will be addressed in upcoming work \cite{Bagchiprep}.}. 
The above equations can be solved to get 
			\begin{equation}
	 \dot{\tau} = \pm P\,,  \quad \dot{x} = \frac{P \tau^2}{\htp^2(\tau^2 - \tau^2_0)}\,,  \quad  \dot{\phi} = \frac{P \tau_0}{\htp (\tau^2 - \tau_0^2)}\,,
			   \label{dotxFSC2}
			\end{equation}
permitting to express $x$ and $\phi$ as functions of $\tau$ 
                \begin{eqnarray}\label{FSCdtdrrot}
                    dx = \pm \frac{d\tau}{f(\tau)}\,,  \qquad
                    d\phi = \pm \frac{N_{\phi}(\tau) d\tau}{f(\tau)}\,.
                \end{eqnarray}
Here $+$ corresponds to right-moving and anti-clockwise rotating rays with respect to $x$ and $\phi$, respectively (and $-$ is defined similarly). The separation of comoving spatial coordinates between two events $(x_i,\tau_i, \phi_i)$ and $(x_j,\tau_j, \phi_j)$ along a null geodesic is got by integrating \eqref{FSCdtdrrot}
                \begin{align}
&x_j - x_i = \pm H(\tau_i, \tau_j)\,, \,
                    \phi_j - \phi_i = \pm G(\tau_i, \tau_j)\,, \label{FSC tphiAB} \\
&\text{where} \,\, H= \int\limits_{\tau_i}^{\tau_j} \frac{\tau^2 d\tau}{\htp^2 (\tau^2 - \tau_0^2)}\,, 
\, G= \frac{\tau_0}{\htp} \int\limits_{\tau_i}^{\tau_j} \frac{d\tau}{(\tau^2 - \tau_0^2)}.\label{eq:integrals}
\end{align}

{\em{Spatial shifts and cosmological chaos.}} To holographically compute the Lyapunov exponent, we consider a left-moving, massless probe that is sent out from past null infinity ($\scri^-$) and crosses the cosmological horizon of the contracting region, adding some tiny amount of energy $\delta M \ll M$ and angular momentum $ \delta J \ll J$ to the background geometry. The probe gets reflected at the timelike causal singularity $\tau=0$ (see \cite{Cornalba:2003kd} for details) and enters the expanding region from the past cosmological horizon. 

\begin{figure}[t]
		            \centering
		            \includegraphics[width=8.5cm]{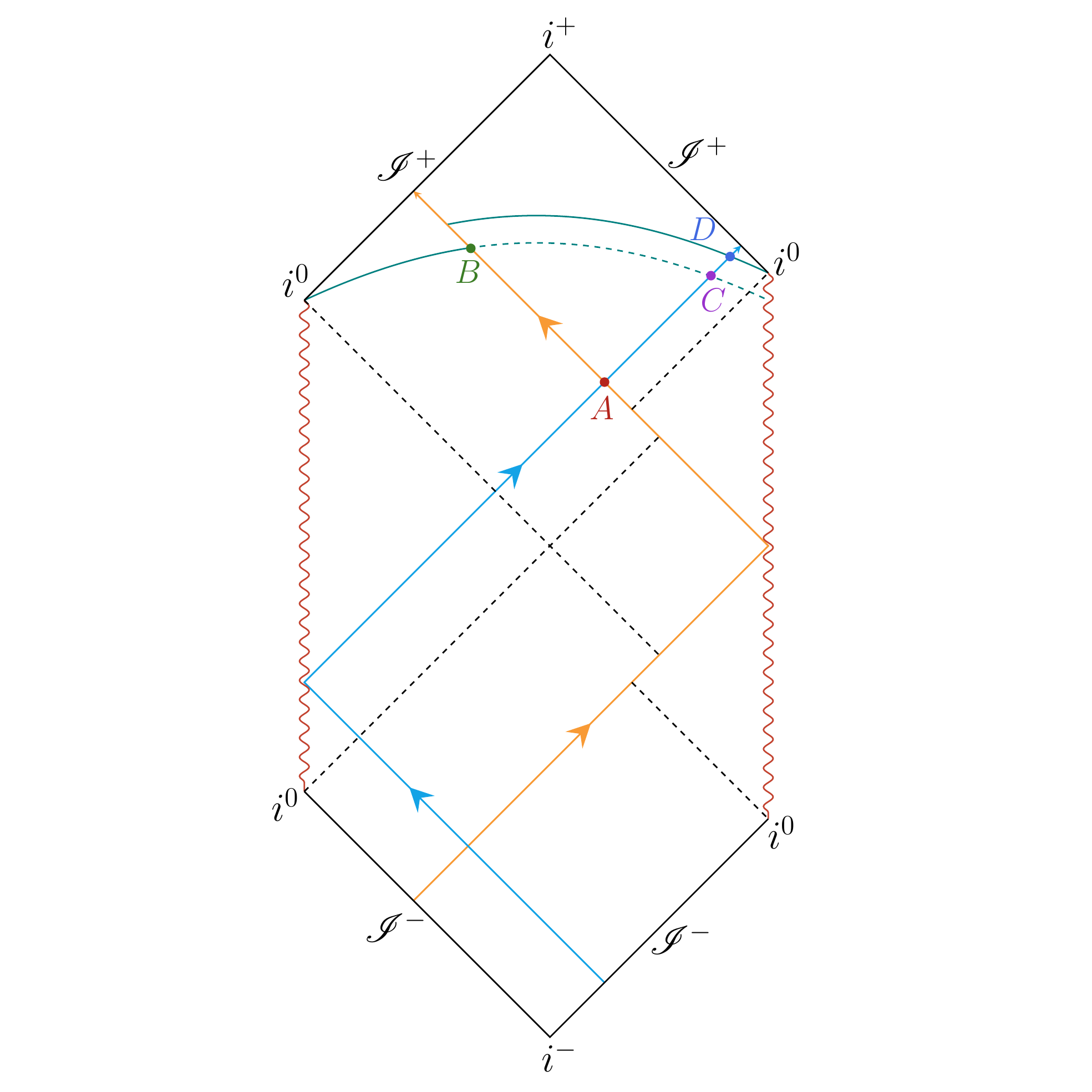}
		            \caption{Penrose diagram depicting the shock wave FSC geometry. Here the lower (upper) part of the diagram corresponds to a contracting (expanding) universe. The {\color{ BurntOrange!80}probe} is sent out from $\scri^-$ and after a reflection at the singularity (wiggly line) emerges from the cosmological horizon (dashed line) where it intersects with the {\color{Cerulean}signal} very close to the horizon at point {\color{BrickRed}$A$} and with the surface {\color{teal}$\tau=\tau_c$} at point {\color{OliveGreen}$B$}. Without taking backreactions into account the {\color{Cerulean}signal} intersects {\color{teal} the surface $\tau=\tau_c$} at the point {\color{DarkOrchid}$C$}. Taking backreactions into account the observer $O'$ gets shifted to the point {\color{RoyalBlue}$D$} in order to still be able to receive the signal at {\color{teal}$\tau=\tau_c$}.}
		            \label{fig:PenroseDiagram}
	                \end{figure}

This is pictorially described in Figure~\ref{fig:PenroseDiagram}. 
An observer $O$ located in the expanding region of the FSC receives this left-moving signal at a time $\tau_c$ and at a point $B=(x_1,\tau_c,\phi_1)$ and eventually sees it moving towards future null infinity ($\scri^+$). Due to the infinite blue shift at the past cosmological horizon the perturbation $(\delta M, \delta J)$ effectively causes the formation of a shock-wave geometry in an FSC background. A right-moving null signal crosses the worldline of the probe very close to the horizon at a point $A=(x_\epsilon,\tau_\epsilon,\phi_\epsilon)$, where $\tau_\epsilon=\tau_0(1+\epsilon)$  for $1\gg\epsilon>0$, and intersects the $\tau=\tau_c$ surface at $C=(x_2,\tau_c,\phi_2)$ in the unperturbed geometry where we place a second observer $O'$. Due to the backreaction of the probe this observer gets shifted to a different spatial point $({\tilde{x}}_2, {\tilde{\phi}}_2)$ on the constant $\tau_c$ surface. Below we show that the comoving distance by which the observer $O'$ is shifted due to the backreaction of the shock wave grows exponentially with the initial comoving distance between the observers $O$ and $O'$, with a numerical prefactor corresponding to the holographic Lyapunov exponent.
                
Applying \eqref{FSC tphiAB} to the spatial separation of comoving coordinates between the events $A$ and $B$ [$x_1-x_\epsilon=-H(\tau_\epsilon,\tau_c)$, $\phi_1-\phi_\epsilon=-G(\tau_\epsilon,\tau_c)$] and between the events $A$ and $C$ [$x_2-x_\epsilon=H(\tau_\epsilon,\tau_c)$, $\phi_2-\phi_\epsilon=G(\tau_\epsilon,\tau_c)$] yields
                \begin{equation}
                    \Delta x = x_2 - x_1 = 2H(\tau_\epsilon, \tau_c), \qquad\;
                    \Delta \phi = 2G(\tau_\epsilon, \tau_c)\,.
                \end{equation}
Solving the integrals \eqref{eq:integrals} for $e^{-\tau_c/\tau_0}\gg\epsilon>\delta\tau_0/\tau_0$ and $\tau_c\gg\tau_0$ gives the leading order behavior of the separation
                \begin{equation}\label{FSCt2-t1}
                    \Delta x = \frac{\beta}{2\pi}\log{\frac{1}{\epsilon}}+ \ldots\,,  \qquad 
                    \Delta\phi = \frac{\mu\beta}{2\pi}\log{\frac{1}{\epsilon}}+ \ldots \,,
                    \end{equation}
expressed in terms of inverse Hawking temperature \mbox{$\beta=2\pi\tau_0/\htp^2$} and angular velocity $\mu=\htp/\tau_0$. To leading order, $\Delta\phi=\mu\Delta x$ and therefore, in what follows we display only $\Delta x$. 

A key aspect of \eqref{FSCt2-t1} is that the spatial intervals are dominated by the red-shift factor $\epsilon$. Holographically, chaos is a consequence of this red-shift factor and backreaction effects from the shock-wave, which together produce an exponential dependence on the separation $\Delta x$, the prefactor of which is the Lyapunov exponent. To determine this exponent, we consider backreactions.

{\em{Backreaction and holographic Lyapunov exponent.}} Backreactions of the shock wave on the FSC background are included by modifying the original FSC parameters $\massp$ and $\jp$ to $\tilde{\massp}=\massp+\delta\massp$ and $\tilde{\jp}=\jp+\delta\jp$. As a consequence, the location of the event horizon also changes to $\tau_0\rightarrow\tau_0+\delta\tau_0$ where $\delta\tau_0=\frac{\beta}{2\pi}\,(\mu\,\delta\jp-\frac12\,\delta\massp)$. This is the first law of thermodynamics for the FSC geometry, 
 \begin{equation}\label{eq:FSCFirstLaw}
                    -T\, \delta S = \delta\mass - \mu\,\delta\angmom\,, \qquad \delta S:=\frac{\delta(\textrm{area})}{4G}=\frac{\pi \delta \tau_0}{2 G} \,,
                \end{equation}
featuring a well-documented sign~\cite{Bagchi:2012xr,Bagchi:2013lma}. The entropy change ${\delta}S$ concurs with the Bekenstein--Hawking area law. 

Due to the backreaction of the shock wave on the FSC geometry the spatial interval $\Delta x = x_j - x_i$ changes  
    			\begin{equation}\label{tildeta-tbH}
    			    \tilde{x}_j - \tilde{x}_i = \pm \int\limits_{\tau_i}^{\tau_j} \frac{d\tau}{\tilde{f}(\tau)} := \pm \tilde{H}(\tau_i,\tau_j), 
    			\end{equation}
Here $\tilde{f}(\tau) = f(\tau)|_{\massp\rightarrow \tilde{\massp},\jp\rightarrow\tilde{\jp}}$.
To linear order in $(\delta\massp, \delta\jp)$  
    			\begin{equation}\label{Htildepurt}
    			    \tilde{H}(\tau_i,\tau_j) = H - \delta\massp \int\limits_{\tau_i}^{\tau_j} \frac{d\tau}{f(\tau)^2} + 2\jp\,\delta\jp  \int\limits_{\tau_i}^{\tau_j} \frac{d\tau}{\tau^2 f(\tau)^2}\,.
    			\end{equation}
The above is valid for general probes carrying energy and angular momentum. Probes without angular momentum, i.e., null geodesics with $L=0$ don't change the total angular momentum of the FSC and thus we have $\delta\jp=0$ and an expanding cosmological horizon in the expanding universe as depicted in Figure~\ref{fig:PenroseDiagram}.
    			
The results \eqref{tildeta-tbH}, \eqref{Htildepurt} yield the interval
    		\begin{equation}\label{eq:ShiftedSeparation}
        			    \tilde{x}_2 - x_\epsilon = \tilde{H}(\tau_\epsilon, \tau_c)
        			    = x_2 - x_\epsilon - \delta\massp \int\limits_{\tau_\epsilon}^{\tau_c} \frac{d\tau}{f(\tau)^2}\,.
    			\end{equation}
Shifting $x_2 \to{\tilde{x}}_2$ takes care of the necessary shift of $O'$ so that the signal is still received at $\tau_c$. The crossing point $x_{\epsilon}$ remains the same as before since the correction to this point is subleading in perturbation parameters.
Evaluating \eqref{eq:ShiftedSeparation}, the leading order contribution to the spatial shift of the observer,
    			\begin{equation}\label{eq:HolographicOSOC}
        			    \tilde{x}_2 - x_2 
        			    \approx \frac{\beta \delta \tau_0}{4\pi \tau_0 \epsilon}=\frac{\beta}{4 \pi}\, (\delta \log{S}) \, e^{\frac{2\pi}{\beta}(x_2 - x_1)}\,,
    			\end{equation}
contains the entropy change $\delta S$ given in \eqref{eq:FSCFirstLaw}. The results above can also be expressed in a coordinate-independent way in terms of proper distances \cite{Bagchiprep}, but we choose a simplified route to match with our field theory answers. 

The expression \eqref{eq:HolographicOSOC} has the same functional form as the amplitude \eqref{eq:IdentityBlockLypunovDecay} computed previously using field theory methods. To connect these results we first take $\tau_c$ to be large, thereby placing the observers $O$ and $O'$ close to $\scri^+$ where the holographic CCFT is encoded. Since \eqref{eq:HolographicOSOC} does not explicitly depend on $\tau_c$, this expression also holds close to $\scri^+$. The inverse Hawking temperature $\beta$ can also be interpreted as the periodicity of the Euclidean time circle and hence the (inverse) temperature that an observer in the dual field theory measures. Thus, we find  the holographic Lyapunov exponent
\beq
\lambda_L = \frac{2\pi}{\beta}\,,
\eeq
in precise agreement with \eqref{eq:whydidthishavenolabel}.

    			%

	{\em{Outlook.}} Our results in this work have moved us into uncharted territories. We generalized 2d CFT results of chaos to non-Lorentzian settings. By studying spatial displacements in cosmological shock-wave solutions, we obtained a holographic Lyapunov exponent matching our field theory results. We shall address further aspects and generalizations of this work in \cite{Bagchiprep}. We close with an important future direction. Chaos is related to the spreading of entanglement \cite{Mezei:2016wfz,Mezei:2016zxg}. 
	Entanglement in Galilean and Carrollian CFTs was studied extensively in \cite{Bagchi:2014iea, Jiang:2017ecm, Hijano:2017eii, Apolo:2020qjm, Grumiller:2019xna,Godet:2019wje}. We would like to understand how our results can help clarify the spread of entanglement and the speed of entanglement (dubbed `butterfly velocity') in these non-Lorentzian theories.

{\em{Acknowledgements.}}
	We thank Jordan Cotler, Diptarka Das, Shahar Hadar, Elizabeth Himwich, Kristan Jensen, Daniel Kapec, Alexandru Lupsasca, Rohan Poojary, Stefan Prohazka, Jakob Salzer, Joan Sim\'on, Douglas Stanford, and Andy Strominger for delightful scientific discussions and comments.
	
	AB is partially supported by a Swarnajayanti Fellowship of the Department of Science and Technology (DST) and the Science and Engineering Research Board (SERB), India. DG is supported by the Austrian Science Fund (FWF), projects P~30822, P~32581, and P~33789.
	The research of MR is supported by the European Union’s Horizon 2020 research and innovation programme under the Marie Skłodowska-Curie grant agreement No.~832542 as well as the DOE grant de-sc/0007870. BR was supported in part by the INSPIRE scholarship of the Department of Science and Technology, Government of India, and the Long Term Visiting Students Program of the International Centre for Theoretical Sciences Bangalore. SC is partially supported by the ISIRD grant 9-252/2016/IITRPR/708.

	\bibliography{Bibli} 

\providecommand{\href}[2]{#2}\begingroup\raggedright\begin{thebibliography}{10}

\bibitem{Lorenz:1972}
E.~Lorenz, ``Predictibility.'' Talk presented at the 139th AAAS meeting, 1972.

\bibitem{Bagchi:2009my}
A.~Bagchi and R.~Gopakumar, ``{Galilean Conformal Algebras and AdS/CFT},'' {\em
  JHEP} {\bf 07} (2009) 037, \href{http://www.arXiv.org/abs/0902.1385}{{\tt
  0902.1385}}.

\bibitem{Bacry:1968zf}
H.~Bacry and J.~Levy-Leblond, ``{Possible kinematics},'' {\em J. Math. Phys.}
  {\bf 9} (1968) 1605--1614.

\bibitem{diFrancesco}
P.~Di~Francesco, P.~Mathieu, and D.~Senechal, {\em Conformal Field Theory}.
\newblock Springer, 1997.

\bibitem{Bagchi:2017yvj}
A.~Bagchi, J.~Chakrabortty, and A.~Mehra, ``{Galilean Field Theories and
  Conformal Structure},'' {\em JHEP} {\bf 04} (2018) 144,
  \href{http://www.arXiv.org/abs/1712.05631}{{\tt 1712.05631}}.

\bibitem{Bagchi:2009pe}
A.~Bagchi, R.~Gopakumar, I.~Mandal, and A.~Miwa, ``{GCA in 2d},'' {\em JHEP}
  {\bf 08} (2010) 004, \href{http://www.arXiv.org/abs/0912.1090}{{\tt
  0912.1090}}.

\bibitem{Roberts:2014ifa}
D.~A. Roberts and D.~Stanford, ``{Two-dimensional conformal field theory and
  the butterfly effect},'' {\em Phys. Rev. Lett.} {\bf 115} (2015), no.~13,
  131603, \href{http://www.arXiv.org/abs/1412.5123}{{\tt 1412.5123}}.

\bibitem{Fitzpatrick:2015zha}
A.~L. Fitzpatrick, J.~Kaplan, and M.~T. Walters, ``{Virasoro Conformal Blocks
  and Thermality from Classical Background Fields},'' {\em JHEP} {\bf 11}
  (2015) 200, \href{http://www.arXiv.org/abs/1501.05315}{{\tt 1501.05315}}.

\bibitem{Perlmutter:2016pkf}
E.~Perlmutter, ``{Bounding the Space of Holographic CFTs with Chaos},'' {\em
  JHEP} {\bf 10} (2016) 069, \href{http://www.arXiv.org/abs/1602.08272}{{\tt
  1602.08272}}.

\bibitem{Bagchi:2012cy}
A.~Bagchi and R.~Fareghbal, ``{BMS/GCA Redux: Towards Flatspace Holography from
  Non-Relativistic Symmetries},'' {\em JHEP} {\bf 10} (2012) 092,
  \href{http://www.arXiv.org/abs/1203.5795}{{\tt 1203.5795}}.

\bibitem{Duval:2014uoa}
C.~Duval, G.~W. Gibbons, P.~A. Horvathy, and P.~M. Zhang, ``{Carroll versus
  Newton and Galilei: two dual non-Einsteinian concepts of time},'' {\em Class.
  Quant. Grav.} {\bf 31} (2014) 085016,
  \href{http://www.arXiv.org/abs/1402.0657}{{\tt 1402.0657}}.

\bibitem{Bagchi:2016bcd}
A.~Bagchi, R.~Basu, A.~Kakkar, and A.~Mehra, ``{Flat Holography: Aspects of the
  dual field theory},'' {\em JHEP} {\bf 12} (2016) 147,
  \href{http://www.arXiv.org/abs/1609.06203}{{\tt 1609.06203}}.

\bibitem{Bagchi:2019clu}
A.~Bagchi, R.~Basu, A.~Mehra, and P.~Nandi, ``{Field Theories on Null
  Manifolds},'' {\em JHEP} {\bf 02} (2020) 141,
  \href{http://www.arXiv.org/abs/1912.09388}{{\tt 1912.09388}}.

\bibitem{Bagchi:2010zz}
A.~Bagchi, ``{Correspondence between Asymptotically Flat Spacetimes and
  Nonrelativistic Conformal Field Theories},'' {\em Phys. Rev. Lett.} {\bf 105}
  (2010) 171601, \href{http://www.arXiv.org/abs/1006.3354}{{\tt 1006.3354}}.

\bibitem{Cornalba:2002fi}
L.~Cornalba and M.~S. Costa, ``{A New cosmological scenario in string
  theory},'' {\em Phys. Rev. D} {\bf 66} (2002) 066001,
  \href{http://www.arXiv.org/abs/hep-th/0203031}{{\tt hep-th/0203031}}.

\bibitem{Cornalba:2003kd}
L.~Cornalba and M.~S. Costa, ``{Time dependent orbifolds and string
  cosmology},'' {\em Fortsch. Phys.} {\bf 52} (2004) 145--199,
  \href{http://www.arXiv.org/abs/hep-th/0310099}{{\tt hep-th/0310099}}.

\bibitem{Shenker:2013pqa}
S.~H. Shenker and D.~Stanford, ``{Black holes and the butterfly effect},'' {\em
  JHEP} {\bf 03} (2014) 067, \href{http://www.arXiv.org/abs/1306.0622}{{\tt
  1306.0622}}.

\bibitem{Shenker:2013yza}
S.~H. Shenker and D.~Stanford, ``{Multiple Shocks},'' {\em JHEP} {\bf 12}
  (2014) 046, \href{http://www.arXiv.org/abs/1312.3296}{{\tt 1312.3296}}.

\bibitem{Maldacena:2015waa}
J.~Maldacena, S.~H. Shenker, and D.~Stanford, ``{A bound on chaos},'' {\em
  JHEP} {\bf 08} (2016) 106, \href{http://www.arXiv.org/abs/1503.01409}{{\tt
  1503.01409}}.

\bibitem{Larkin:1969aa}
A.~Larkin and Y.~Ovchinnikov, ``Quasiclassical method in the theory of
  superconductivity,'' {\em Sov.Phys.JETP} {\bf 28} (19669) 1200.

\bibitem{Roberts:2014isa}
D.~A. Roberts, D.~Stanford, and L.~Susskind, ``{Localized shocks},'' {\em JHEP}
  {\bf 03} (2015) 051, \href{http://www.arXiv.org/abs/1409.8180}{{\tt
  1409.8180}}.

\bibitem{Shenker:2014cwa}
S.~H. Shenker and D.~Stanford, ``{Stringy effects in scrambling},'' {\em JHEP}
  {\bf 05} (2015) 132, \href{http://www.arXiv.org/abs/1412.6087}{{\tt
  1412.6087}}.

\bibitem{Kitaev:2014aa}
A.~Kitaev, ``{Hidden Correlations in the Hawking Radiation and Thermal
  Noise},'' {\em talk given at KITP, Santa Barbara} (2014).
  \href{http://online.kitp.ucsb.edu/online/joint98/kitaev/}{http://online.kitp.ucsb.edu/online/joint98/kitaev/}.

\bibitem{Haehl:2017qfl}
F.~M. Haehl, R.~Loganayagam, P.~Narayan, and M.~Rangamani, ``{Classification of
  out-of-time-order correlators},'' {\em SciPost Phys.} {\bf 6} (2019), no.~1,
  001, \href{http://www.arXiv.org/abs/1701.02820}{{\tt 1701.02820}}.

\bibitem{Note1}
The approximation of large central charge corresponds to the classical
  approximation on the gravity side (small Newton constant) \cite
  {Barnich:2006av}, whenever there is a gravity dual. Therefore, this
  approximation is not only convenient, it is also pertinent to the holographic
  calculation in the second half of our work.

\bibitem{Bagchi:2016geg}
A.~Bagchi, M.~Gary, and Zodinmawia, ``{Bondi-Metzner-Sachs bootstrap},'' {\em
  Phys. Rev. D} {\bf 96} (2017), no.~2, 025007,
  \href{http://www.arXiv.org/abs/1612.01730}{{\tt 1612.01730}}.

\bibitem{Bagchi:2017cpu}
A.~Bagchi, M.~Gary, and Zodinmawia, ``{The nuts and bolts of the BMS
  Bootstrap},'' {\em Class. Quant. Grav.} {\bf 34} (2017), no.~17, 174002,
  \href{http://www.arXiv.org/abs/1705.05890}{{\tt 1705.05890}}.

\bibitem{Hijano:2017eii}
E.~Hijano and C.~Rabideau, ``{Holographic entanglement and Poincar\'e blocks in
  three-dimensional flat space},'' {\em JHEP} {\bf 05} (2018) 068,
  \href{http://www.arXiv.org/abs/1712.07131}{{\tt 1712.07131}}.

\bibitem{Hijano:2018nhq}
E.~Hijano, ``{Semi-classical BMS$_{3}$ blocks and flat holography},'' {\em
  JHEP} {\bf 10} (2018) 044, \href{http://www.arXiv.org/abs/1805.00949}{{\tt
  1805.00949}}.

\bibitem{Merbis:2019wgk}
W.~Merbis and M.~Riegler, ``{Geometric actions and flat space holography},''
  {\em JHEP} {\bf 02} (2020) 125,
  \href{http://www.arXiv.org/abs/1912.08207}{{\tt 1912.08207}}.

\bibitem{Note2}
Since $c_M$ has a physical dimension a more precise definition of scrambling
  time is $t_\ast \sim \protect \frac {\beta }{2\pi }\protect \qopname \relax
  o{log}\protect \frac {c_M}{\xi _n}$, where $\xi _n$ is either $\xi _V$ or
  $\xi _W$, depending on which weight is bigger. However, since we assume the
  weights to be finite and $c_M/\xi _n$ to be large, the quantity $c_M$ must be
  large and hence dominates the scrambling time.

\bibitem{Note3}
This result can be extended to the case where $c_L\protect \neq 0$ since a
  non-zero $c_L$ only affects the $\protect \mathcal {O}(1/c_M^2)$ terms.

\bibitem{Maldacena:1997re}
J.~M. Maldacena, ``{The Large N limit of superconformal field theories and
  supergravity},'' {\em Adv. Theor. Math. Phys.} {\bf 2} (1998) 231--252,
  \href{http://www.arXiv.org/abs/hep-th/9711200}{{\tt hep-th/9711200}}.

\bibitem{Barnich:2012aw}
G.~Barnich, A.~Gomberoff, and H.~A. Gonzalez, ``{The Flat limit of three
  dimensional asymptotically anti-de Sitter spacetimes},'' {\em Phys. Rev. D}
  {\bf 86} (2012) 024020, \href{http://www.arXiv.org/abs/1204.3288}{{\tt
  1204.3288}}.

\bibitem{Bagchi:2012yk}
A.~Bagchi, S.~Detournay, and D.~Grumiller, ``{Flat-Space Chiral Gravity},''
  {\em Phys. Rev. Lett.} {\bf 109} (2012) 151301,
  \href{http://www.arXiv.org/abs/1208.1658}{{\tt 1208.1658}}.

\bibitem{Bagchi:2012xr}
A.~Bagchi, S.~Detournay, R.~Fareghbal, and J.~Sim\'on, ``{Holography of 3D Flat
  Cosmological Horizons},'' {\em Phys. Rev. Lett.} {\bf 110} (2013), no.~14,
  141302, \href{http://www.arXiv.org/abs/1208.4372}{{\tt 1208.4372}}.

\bibitem{Barnich:2012xq}
G.~Barnich, ``{Entropy of three-dimensional asymptotically flat cosmological
  solutions},'' {\em JHEP} {\bf 10} (2012) 095,
  \href{http://www.arXiv.org/abs/1208.4371}{{\tt 1208.4371}}.

\bibitem{Barnich:2012rz}
G.~Barnich, A.~Gomberoff, and H.~A. Gonz\'alez, ``{Three-dimensional
  Bondi-Metzner-Sachs invariant two-dimensional field theories as the flat
  limit of Liouville theory},'' {\em Phys. Rev. D} {\bf 87} (2013), no.~12,
  124032, \href{http://www.arXiv.org/abs/1210.0731}{{\tt 1210.0731}}.

\bibitem{Bagchi:2013lma}
A.~Bagchi, S.~Detournay, D.~Grumiller, and J.~Simon, ``{Cosmic Evolution from
  Phase Transition of Three-Dimensional Flat Space},'' {\em Phys. Rev. Lett.}
  {\bf 111} (2013), no.~18, 181301,
  \href{http://www.arXiv.org/abs/1305.2919}{{\tt 1305.2919}}.

\bibitem{Afshar:2013vka}
H.~Afshar, A.~Bagchi, R.~Fareghbal, D.~Grumiller, and J.~Rosseel, ``{Spin-3
  Gravity in Three-Dimensional Flat Space},'' {\em Phys. Rev. Lett.} {\bf 111}
  (2013), no.~12, 121603, \href{http://www.arXiv.org/abs/1307.4768}{{\tt
  1307.4768}}.

\bibitem{Gonzalez:2013oaa}
H.~A. Gonzalez, J.~Matulich, M.~Pino, and R.~Troncoso, ``{Asymptotically flat
  spacetimes in three-dimensional higher spin gravity},'' {\em JHEP} {\bf 09}
  (2013) 016, \href{http://www.arXiv.org/abs/1307.5651}{{\tt 1307.5651}}.

\bibitem{Fareghbal:2013ifa}
R.~Fareghbal and A.~Naseh, ``{Flat-Space Energy-Momentum Tensor from BMS/GCA
  Correspondence},'' {\em JHEP} {\bf 03} (2014) 005,
  \href{http://www.arXiv.org/abs/1312.2109}{{\tt 1312.2109}}.

\bibitem{Detournay:2014fva}
S.~Detournay, D.~Grumiller, F.~Sch\"oller, and J.~Sim\'on, ``{Variational
  principle and one-point functions in three-dimensional flat space Einstein
  gravity},'' {\em Phys. Rev. D} {\bf 89} (2014), no.~8, 084061,
  \href{http://www.arXiv.org/abs/1402.3687}{{\tt 1402.3687}}.

\bibitem{Bagchi:2014iea}
A.~Bagchi, R.~Basu, D.~Grumiller, and M.~Riegler, ``{Entanglement entropy in
  Galilean conformal field theories and flat holography},'' {\em Phys. Rev.
  Lett.} {\bf 114} (2015), no.~11, 111602,
  \href{http://www.arXiv.org/abs/1410.4089}{{\tt 1410.4089}}.

\bibitem{Bagchi:2015wna}
A.~Bagchi, D.~Grumiller, and W.~Merbis, ``{Stress tensor correlators in
  three-dimensional gravity},'' {\em Phys. Rev. D} {\bf 93} (2016), no.~6,
  061502, \href{http://www.arXiv.org/abs/1507.05620}{{\tt 1507.05620}}.

\bibitem{Hartong:2015usd}
J.~Hartong, ``{Holographic Reconstruction of 3D Flat Space-Time},'' {\em JHEP}
  {\bf 10} (2016) 104, \href{http://www.arXiv.org/abs/1511.01387}{{\tt
  1511.01387}}.

\bibitem{Barnich:2006av}
G.~Barnich and G.~Compere, ``{Classical central extension for asymptotic
  symmetries at null infinity in three spacetime dimensions},'' {\em Class.
  Quant. Grav.} {\bf 24} (2007) F15--F23,
  \href{http://www.arXiv.org/abs/gr-qc/0610130}{{\tt gr-qc/0610130}}.

\bibitem{Banados:1992wn}
M.~Ba\~nados, C.~Teitelboim, and J.~Zanelli, ``{The Black hole in
  three-dimensional space-time},'' {\em Phys. Rev. Lett.} {\bf 69} (1992)
  1849--1851, \href{http://www.arXiv.org/abs/hep-th/9204099}{{\tt
  hep-th/9204099}}.

\bibitem{Banados:1992gq}
M.~Ba\~nados, M.~Henneaux, C.~Teitelboim, and J.~Zanelli, ``{Geometry of the
  (2+1) black hole},'' {\em Phys. Rev. D} {\bf 48} (1993) 1506--1525,
  \href{http://www.arXiv.org/abs/gr-qc/9302012}{{\tt gr-qc/9302012}}. [Erratum:
  Phys.Rev.D 88, 069902 (2013)].

\bibitem{Engelsoy:2016xyb}
J.~Engels\"oy, T.~G. Mertens, and H.~Verlinde, ``{An investigation of AdS$_{2}$
  backreaction and holography},'' {\em JHEP} {\bf 07} (2016) 139,
  \href{http://www.arXiv.org/abs/1606.03438}{{\tt 1606.03438}}.

\bibitem{Grumiller:2020fbb}
D.~Grumiller and R.~McNees, ``{Universal flow equations and chaos bound
  saturation in 2d dilaton gravity},'' {\em JHEP} {\bf 01} (2021) 112,
  \href{http://www.arXiv.org/abs/2007.03673}{{\tt 2007.03673}}.

\bibitem{Note4}
The case $L\protect \neq 0$ yields essentially the same results and will be
  addressed in upcoming work \cite {Bagchiprep}.

\bibitem{Bagchiprep}
A.~Bagchi, S.~Chakrabortty, D.~Grumiller, B.~Radhakrishnan, M.~Riegler, and
  A.~Sinha, ``Chaos and holography in {BMS} invariant field theories.'' in
  preparation.

\bibitem{Mezei:2016wfz}
M.~Mezei and D.~Stanford, ``{On entanglement spreading in chaotic systems},''
  {\em JHEP} {\bf 05} (2017) 065,
  \href{http://www.arXiv.org/abs/1608.05101}{{\tt 1608.05101}}.

\bibitem{Mezei:2016zxg}
M.~Mezei, ``{On entanglement spreading from holography},'' {\em JHEP} {\bf 05}
  (2017) 064, \href{http://www.arXiv.org/abs/1612.00082}{{\tt 1612.00082}}.

\bibitem{Jiang:2017ecm}
H.~Jiang, W.~Song, and Q.~Wen, ``{Entanglement Entropy in Flat Holography},''
  {\em JHEP} {\bf 07} (2017) 142,
  \href{http://www.arXiv.org/abs/1706.07552}{{\tt 1706.07552}}.

\bibitem{Apolo:2020qjm}
L.~Apolo, H.~Jiang, W.~Song, and Y.~Zhong, ``{Modular Hamiltonians in flat
  holography and (W)AdS/WCFT},'' {\em JHEP} {\bf 09} (2020) 033,
  \href{http://www.arXiv.org/abs/2006.10741}{{\tt 2006.10741}}.

\bibitem{Grumiller:2019xna}
D.~Grumiller, P.~Parekh, and M.~Riegler, ``{Local quantum energy conditions in
  non-Lorentz-invariant quantum field theories},'' {\em Phys. Rev. Lett.} {\bf
  123} (2019), no.~12, 121602, \href{http://www.arXiv.org/abs/1907.06650}{{\tt
  1907.06650}}.

\bibitem{Godet:2019wje}
V.~Godet and C.~Marteau, ``{Gravitation in flat spacetime from entanglement},''
  {\em JHEP} {\bf 12} (2019) 057,
  \href{http://www.arXiv.org/abs/1908.02044}{{\tt 1908.02044}}.

\end{thebibliography}\endgroup
    \bibliographystyle{fullsort}

	\end{document}